

%
%
%
\def\unredoffs{} \def\redoffs{\voffset=-.31truein\hoffset=-.59truein}
\def\speclscape{\special{ps: landscape}}
%
%
%
%
\newbox\leftpage \newdimen\fullhsize \newdimen\hstitle \newdimen\hsbody
\tolerance=1000\hfuzz=2pt
\catcode`\@=11 
\def\bigans{b }
\def\answ{b }

%
\ifx\answ\bigans\message{(This will come out unreduced.}
\magnification=1200\unredoffs\baselineskip=16pt plus 2pt minus 1pt
\hsbody=\hsize \hstitle=\hsize 
\else\message{(This will be reduced.} \let\l@r=L
\magnification=1000\baselineskip=16pt plus 2pt minus 1pt \vsize=7truein
\redoffs \hstitle=8truein\hsbody=4.75truein\fullhsize=10truein\hsize=\hsbody
\output={\ifnum\pageno=0 
  \shipout\vbox{\speclscape{\hsize\fullhsize\makeheadline}
    \hbox to \fullhsize{\hfill\pagebody\hfill}}\advancepageno
  \else
  \almostshipout{\leftline{\vbox{\pagebody\makefootline}}}\advancepageno
  \fi}
\def\almostshipout#1{\if L\l@r \count1=1 \message{[\the\count0.\the\count1]}
      \global\setbox\leftpage=#1 \global\let\l@r=R
 \else \count1=2
  \shipout\vbox{\speclscape{\hsize\fullhsize\makeheadline}
      \hbox to\fullhsize{\box\leftpage\hfil#1}}  \global\let\l@r=L\fi}
\fi
%
\newcount\yearltd\yearltd=\year\advance\yearltd by -1900

\def\Title#1#2{\nopagenumbers\abstractfont\hsize=\hstitle\rightline{#1}%
\vskip 1in\centerline{\titlefont #2}\abstractfont\vskip .5in\pageno=0}
\def\Date#1{\vfill\leftline{#1}\tenpoint\supereject\global\hsize=\hsbody%
\footline={\hss\tenrm\folio\hss}}
%

\def\draftmode{\message{ DRAFTMODE }\def\draftdate{{\rm preliminary draft:
\number\month/\number\day/\number\yearltd\ \ \hourmin}}%
\headline={\hfil\draftdate}\writelabels\baselineskip=20pt plus 2pt minus 2pt
 {\count255=\time\divide\count255 by 60 \xdef\hourmin{\number\count255}
  \multiply\count255 by-60\advance\count255 by\time
  \xdef\hourmin{\hourmin:\ifnum\count255<10 0\fi\the\count255}}}
\def\nolabels{\def\wrlabeL##1{}\def\eqlabeL##1{}\def\reflabeL##1{}}
\def\writelabels{\def\wrlabeL##1{\leavevmode\vadjust{\rlap{\smash%
{\line{{\escapechar=` \hfill\rlap{\sevenrm\hskip.03in\string##1}}}}}}}%
\def\eqlabeL##1{{\escapechar-1\rlap{\sevenrm\hskip.05in\string##1}}}%
\def\reflabeL##1{\noexpand\llap{\noexpand\sevenrm\string\string\string##1}}}
\nolabels
%
\global\newcount\secno \global\secno=0
\global\newcount\meqno \global\meqno=1
\def\newsec#1{\global\advance\secno by1\message{(\the\secno. #1)}
\global\subsecno=0\eqnres@t\noindent{\bf\the\secno. #1}
\writetoca{{\secsym} {#1}}\par\nobreak\medskip\nobreak}
\def\eqnres@t{\xdef\secsym{\the\secno.}\global\meqno=1\bigbreak\bigskip}
\def\sequentialequations{\def\eqnres@t{\bigbreak}}\xdef\secsym{}
\global\newcount\subsecno \global\subsecno=0
\def\subsec#1{\global\advance\subsecno by1\message{(\secsym\the\subsecno. #1)}
\ifnum\lastpenalty>9000\else\bigbreak\fi
\noindent{\it\secsym\the\subsecno. #1}\writetoca{\string\quad
{\secsym\the\subsecno.} {#1}}\par\nobreak\medskip\nobreak}
\def\appendix#1#2{\global\meqno=1\global\subsecno=0\xdef\secsym{\hbox{#1.}}
\bigbreak\bigskip\noindent{\bf Appendix #1. #2}\message{(#1. #2)}
\writetoca{Appendix {#1.} {#2}}\par\nobreak\medskip\nobreak}
%
%
\def\eqnn#1{\xdef #1{(\secsym\the\meqno)}\writedef{#1\leftbracket#1}%
\global\advance\meqno by1\wrlabeL#1}
\def\eqna#1{\xdef #1##1{\hbox{$(\secsym\the\meqno##1)$}}
\writedef{#1\numbersign1\leftbracket#1{\numbersign1}}%
\global\advance\meqno by1\wrlabeL{#1$\{\}$}}
\def\eqn#1#2{\xdef #1{(\secsym\the\meqno)}\writedef{#1\leftbracket#1}%
\global\advance\meqno by1$$#2\eqno#1\eqlabeL#1$$}
%
\newskip\footskip\footskip14pt plus 1pt minus 1pt 
\def\footnotefont{\ninepoint}\def\f@t#1{\footnotefont #1\@foot}
\def\f@@t{\baselineskip\footskip\bgroup\footnotefont\aftergroup\@foot\let\next}
\setbox\strutbox=\hbox{\vrule height9.5pt depth4.5pt width0pt}
\global\newcount\ftno \global\ftno=0
\def\foot{\global\advance\ftno by1\footnote{$^{\the\ftno}$}}
%
\newwrite\ftfile
\def\footend{\def\foot{\global\advance\ftno by1\chardef\wfile=\ftfile
$^{\the\ftno}$\ifnum\ftno=1\immediate\openout\ftfile=foots.tmp\fi%
\immediate\write\ftfile{\noexpand\smallskip%
\noexpand\item{f\the\ftno:\ }\pctsign}\findarg}%
\def\footatend{\vfill\eject\immediate\closeout\ftfile{\parindent=20pt
\centerline{\bf Footnotes}\nobreak\bigskip\input foots.tmp }}}
\def\footatend{}
%
%
\global\newcount\refno \global\refno=1
\newwrite\rfile
\def\ref{[\the\refno]\nref}
\def\nref#1{\xdef#1{[\the\refno]}\writedef{#1\leftbracket#1}%
\ifnum\refno=1\immediate\openout\rfile=refs.tmp\fi
\global\advance\refno by1\chardef\wfile=\rfile\immediate
\write\rfile{\noexpand\item{#1\ }\reflabeL{#1\hskip.31in}\pctsign}\findarg}
\def\findarg#1#{\begingroup\obeylines\newlinechar=`\^^M\pass@rg}
{\obeylines\gdef\pass@rg#1{\writ@line\relax #1^^M\hbox{}^^M}%
\gdef\writ@line#1^^M{\expandafter\toks0\expandafter{\striprel@x #1}%
\edef\next{\the\toks0}\ifx\next\em@rk\let\next=\endgroup\else\ifx\next\empty%
\else\immediate\write\wfile{\the\toks0}\fi\let\next=\writ@line\fi\next\relax}}
\def\striprel@x#1{} \def\em@rk{\hbox{}}
\def\lref{\begingroup\obeylines\lr@f}
\def\lr@f#1#2{\gdef#1{\ref#1{#2}}\endgroup\unskip}

\def\addref#1{\immediate\write\rfile{\noexpand\item{}#1}} 
\def\startrefs#1{\immediate\openout\rfile=refs.tmp\refno=#1}
\def\xref{\expandafter\xr@f}\def\xr@f[#1]{#1}
\def\refs#1{\count255=1[\r@fs #1{\hbox{}}]}
\def\r@fs#1{\ifx\und@fined#1\message{reflabel \string#1 is undefined.}%
\nref#1{need to supply reference \string#1.}\fi%
\vphantom{\hphantom{#1}}\edef\next{#1}\ifx\next\em@rk\def\next{}%
\else\ifx\next#1\ifodd\count255\relax\xref#1\count255=0\fi%
\else#1\count255=1\fi\let\next=\r@fs\fi\next}
%

%
\newwrite\ffile\global\newcount\figno \global\figno=1
\def\fig{fig.~\the\figno\nfig}
\def\nfig#1{\xdef#1{fig.~\the\figno}%
\writedef{#1\leftbracket fig.\noexpand~\the\figno}%
\ifnum\figno=1\immediate\openout\ffile=figs.tmp\fi\chardef\wfile=\ffile%
\immediate\write\ffile{\noexpand\medskip\noexpand\item{Fig.\ \the\figno. }
\reflabeL{#1\hskip.55in}\pctsign}\global\advance\figno by1\findarg}
\def\xfig{\expandafter\xf@g}\def\xf@g fig.\penalty\@M\ {}
\def\figs#1{figs.~\f@gs #1{\hbox{}}}
\def\f@gs#1{\edef\next{#1}\ifx\next\em@rk\def\next{}\else
\ifx\next#1\xfig #1\else#1\fi\let\next=\f@gs\fi\next}
\newwrite\lfile
{\escapechar-1\xdef\pctsign{\string\%}\xdef\leftbracket{\string\{}
\xdef\rightbracket{\string\}}\xdef\numbersign{\string\#}}

\def\writestop{\def\writestoppt{\immediate\write\lfile{\string\pageno%
\the\pageno\string\startrefs\leftbracket\the\refno\rightbracket%
\string\def\string\secsym\leftbracket\secsym\rightbracket%
\string\secno\the\secno\string\meqno\the\meqno}\immediate\closeout\lfile}}
\def\writestoppt{}\def\writedef#1{}
\def\seclab#1{\xdef #1{\the\secno}\writedef{#1\leftbracket#1}\wrlabeL{#1=#1}}
\def\subseclab#1{\xdef #1{\secsym\the\subsecno}%
\writedef{#1\leftbracket#1}\wrlabeL{#1=#1}}
\newwrite\tfile \def\writetoca#1{}
\def\leaderfill{\leaders\hbox to 1em{\hss.\hss}\hfill}
\def\writetoc{\immediate\openout\tfile=toc.tmp
   \def\writetoca##1{{\edef\next{\write\tfile{\noindent ##1
   \string\leaderfill {\noexpand\number\pageno} \par}}\next}}}

%
\catcode`\@=12 
%
\edef\tfontsize{\ifx\answ\bigans scaled\magstep3\else scaled\magstep4\fi}
\font\titlerm=cmr10 \tfontsize \font\titlerms=cmr7 \tfontsize
\font\titlermss=cmr5 \tfontsize \font\titlei=cmmi10 \tfontsize
\font\titleis=cmmi7 \tfontsize \font\titleiss=cmmi5 \tfontsize
\font\titlesy=cmsy10 \tfontsize \font\titlesys=cmsy7 \tfontsize
\font\titlesyss=cmsy5 \tfontsize \font\titleit=cmti10 \tfontsize
\skewchar\titlei='177 \skewchar\titleis='177 \skewchar\titleiss='177
\skewchar\titlesy='60 \skewchar\titlesys='60 \skewchar\titlesyss='60
\def\titlefont{\def\rm{\fam0\titlerm}
\textfont0=\titlerm \scriptfont0=\titlerms \scriptscriptfont0=\titlermss
\textfont1=\titlei \scriptfont1=\titleis \scriptscriptfont1=\titleiss
\textfont2=\titlesy \scriptfont2=\titlesys \scriptscriptfont2=\titlesyss
\textfont\itfam=\titleit \def\it{\fam\itfam\titleit}\rm}
 \ifx\answ\bigans\else scaled\magstep1\fi
\ifx\answ\bigans\def\abstractfont{\tenpoint}\else
\font\abssl=cmsl10 scaled \magstep1
\font\absrm=cmr10 scaled\magstep1 \font\absrms=cmr7 scaled\magstep1
\font\absrmss=cmr5 scaled\magstep1 \font\absi=cmmi10 scaled\magstep1
\font\absis=cmmi7 scaled\magstep1 \font\absiss=cmmi5 scaled\magstep1
\font\abssy=cmsy10 scaled\magstep1 \font\abssys=cmsy7 scaled\magstep1
\font\abssyss=cmsy5 scaled\magstep1 \font\absbf=cmbx10 scaled\magstep1
\skewchar\absi='177 \skewchar\absis='177 \skewchar\absiss='177
\skewchar\abssy='60 \skewchar\abssys='60 \skewchar\abssyss='60
\def\abstractfont{\def\rm{\fam0\absrm}
\textfont0=\absrm \scriptfont0=\absrms \scriptscriptfont0=\absrmss
\textfont1=\absi \scriptfont1=\absis \scriptscriptfont1=\absiss
\textfont2=\abssy \scriptfont2=\abssys \scriptscriptfont2=\abssyss
\textfont\itfam=\bigit \def\it{\fam\itfam\bigit}\def\footnotefont{\tenpoint}%
\textfont\slfam=\abssl \def\sl{\fam\slfam\abssl}%
\textfont\bffam=\absbf \def\bf{\fam\bffam\absbf}\rm}\fi
\def\tenpoint{\def\rm{\fam0\tenrm}
\textfont0=\tenrm \scriptfont0=\sevenrm \scriptscriptfont0=\fiverm
\textfont1=\teni  \scriptfont1=\seveni  \scriptscriptfont1=\fivei
\textfont2=\tensy \scriptfont2=\sevensy \scriptscriptfont2=\fivesy
\textfont\itfam=\tenit \def\it{\fam\itfam\tenit}\def\footnotefont{\ninepoint}%
\textfont\bffam=\tenbf \def\bf{\fam\bffam\tenbf}\def\sl{\fam\slfam\tensl}\rm}
\font\ninerm=cmr9 \font\sixrm=cmr6 \font\ninei=cmmi9 \font\sixi=cmmi6
\font\ninesy=cmsy9 \font\sixsy=cmsy6 \font\ninebf=cmbx9
\font\nineit=cmti9 \font\ninesl=cmsl9 \skewchar\ninei='177
\skewchar\sixi='177 \skewchar\ninesy='60 \skewchar\sixsy='60
\def\ninepoint{\def\rm{\fam0\ninerm}
\textfont0=\ninerm \scriptfont0=\sixrm \scriptscriptfont0=\fiverm
\textfont1=\ninei \scriptfont1=\sixi \scriptscriptfont1=\fivei
\textfont2=\ninesy \scriptfont2=\sixsy \scriptscriptfont2=\fivesy
\textfont\itfam=\ninei \def\it{\fam\itfam\nineit}\def\sl{\fam\slfam\ninesl}%
\textfont\bffam=\ninebf \def\bf{\fam\bffam\ninebf}\rm}
%
%

\hyphenation{anom-aly anom-alies coun-ter-term coun-ter-terms}
\def\inv{^{\raise.15ex\hbox{${\scriptscriptstyle -}$}\kern-.05em 1}}

\def\Dsl{\,\raise.15ex\hbox{/}\mkern-13.5mu D} 
\def\dsl{\raise.15ex\hbox{/}\kern-.57em\partial}

\def\tr{{\rm tr}} 
\font\bigit=cmti10 scaled \magstep1
\def\lspace{\ifx\answ\bigans{}\else\qquad\fi}
\def\lbspace{\ifx\answ\bigans{}\else\hskip-.2in\fi} 
\def\boxeqn#1{\vcenter{\vbox{\hrule\hbox{\vrule\kern3pt\vbox{\kern3pt
	\hbox{${\displaystyle #1}$}\kern3pt}\kern3pt\vrule}\hrule}}}
\def\mbox#1#2{\vcenter{\hrule \hbox{\vrule height#2in
		\kern#1in \vrule} \hrule}}  
%

\def\darr#1{\raise1.5ex\hbox{$\leftrightarrow$}\mkern-16.5mu #1}

\def\roughly#1{\raise.3ex\hbox{$#1$\kern-.75em\lower1ex\hbox{$\sim$}}}

\let\includefigures=\iftrue
\let\useblackboard=\iftrue
\newfam\black

\includefigures
\message{If you do not have epsf.tex (to include figures),}
\message{change the option at the top of the tex file.}
\input epsf
\def\figin{\epsfcheck\figin}\def\figins{\epsfcheck\figins}
\def\epsfcheck{\ifx\epsfbox\UnDeFiNeD
\message{(NO epsf.tex, FIGURES WILL BE IGNORED)}
\gdef\figin##1{\vskip2in}\gdef\figins##1{\hskip.5in}
\else\message{(FIGURES WILL BE INCLUDED)}%
\gdef\figin##1{##1}\gdef\figins##1{##1}\fi}
\def\DefWarn#1{}
\def\figinsert{\goodbreak\midinsert}
\def\ifig#1#2#3{\DefWarn#1\xdef#1{fig.~\the\figno}
\writedef{#1\leftbracket fig.\noexpand~\the\figno}%
\figinsert\figin{\centerline{#3}}\medskip\centerline{\vbox{
\baselineskip12pt\advance\hsize by -1truein
\noindent\footnotefont{\bf Fig.~\the\figno:} #2}}
\endinsert\global\advance\figno by1}
\else
\def\ifig#1#2#3{\xdef#1{fig.~\the\figno}
\writedef{#1\leftbracket fig.\noexpand~\the\figno}%
\global\advance\figno by1} \fi

\def\AdS{{\bf AdS}}

\def\id{{1 \kern-.28em {\rm l}}}
\def\N{{\cal N}}

\def\K3{{\bf K3}}
\def\journal#1&#2(#3){\unskip, \sl #1\ \bf #2 \rm(19#3) }
\def\andjournal#1&#2(#3){\sl #1~\bf #2 \rm (19#3) }

\def\hat{\widehat}

\def\tilde{\widetilde}

\def\frac#1#2{{#1\over#2}}

\def\inbar{\,\vrule height1.5ex width.4pt depth0pt}
\def\IC{\relax\hbox{$\inbar\kern-.3em{\rm C}$}}
\def\IR{\relax{\rm I\kern-.18em R}}
\def\IP{\relax{\rm I\kern-.18em P}}

%
%

%
\catcode`\@=11
\def\slash#1{\mathord{\mathpalette\c@ncel{#1}}}
\overfullrule=0pt
\def\AA{{\cal A}}
\def\BB{{\cal B}}

\def\FF{{\cal F}}
\def\GG{{\cal G}}

\def\LL{{\cal L}}
\def\NN{{\cal N}}
\def\OO{{\cal O}}

\def\underrel#1\over#2{\mathrel{\mathop{\kern\z@#1}\limits_{#2}}}

\catcode`\@=12


%

\def\det{{\rm det}}
\def\tr{{\rm tr}}

\def\det{{\rm det}}


\def\p{{\partial}}

\def\ra{{\rightarrow}}

\def\tmu{{\tilde \mu}}

\def\Af{{\widetilde A}}
\def\tm{{\tilde{m}}}
\def\tmu{{\tilde{\mu}}}
\def\tq{{\tilde{q}}}
\def\tom{{\tilde{\omega}}}

\def\k{{\bf k}}
\def\NN{{\cal N}}
\def\x{{\xi}}
\def\xf{{\widetilde \xi}}
\def\C{{\tilde C}}
\def\ta{{\tilde a}}
\def\tb{{\tilde b}}

\lref\ErdmengerAP{
  J.~Erdmenger, M.~Kaminski and F.~Rust,
  ``Isospin diffusion in thermal AdS/CFT with flavor,''
  Phys.\ Rev.\  D {\bf 76}, 046001 (2007)
  [arXiv:0704.1290 [hep-th]].
}

\lref\MM{
  D.~Mateos, S.~Matsuura, R.~C.~Myers and R.~M.~Thomson,
  ``Holographic phase transitions at finite chemical potential,''
  arXiv:0709.1225 [hep-th].
}

\lref\NakamuraNX{
  S.~Nakamura, Y.~Seo, S.~J.~Sin and K.~P.~Yogendran,
  ``Baryon-charge Chemical Potential in AdS/CFT,''
  arXiv:0708.2818 [hep-th].
}

\lref\GhorokuRE{
  K.~Ghoroku, M.~Ishihara and A.~Nakamura,
  ``D3/D7 holographic Gauge theory and Chemical potential,''
  arXiv:0708.3706 [hep-th].
}

\lref\KarchBR{
  A.~Karch and A.~O'Bannon,
  ``Holographic Thermodynamics at Finite Baryon Density: Some Exact Results,''
  arXiv:0709.0570 [hep-th].
}

\lref\KarchFA{
  A.~Karch, D.~T.~Son and A.~O.~Starinets,
  ``Zero Sound from Holography,''
  arXiv:0806.3796 [hep-th].
}

\lref\KarchSH{
  A.~Karch and E.~Katz,
  ``Adding flavor to AdS/CFT,''
  JHEP {\bf 0206}, 043 (2002)
  [arXiv:hep-th/0205236].
}

\lref\KovtunEV{
  P.~K.~Kovtun and A.~O.~Starinets,
  ``Quasinormal modes and holography,''
  Phys.\ Rev.\  D {\bf 72}, 086009 (2005)
  [arXiv:hep-th/0506184].
}

\lref\MyersMS{
  R.~C.~Myers and A.~Sinha,
  ``The fast life of holographic mesons,''
  JHEP {\bf 0806}, 052 (2008)
  [arXiv:hep-th/08042423].
}

\lref\MasSG{
  J.~Mas, J.~P.~Shock, J.~Tarrio and D.~Zoakos
  ``Holographic Spectral Functions at Finite Baryon Density,''
   [arXiv:hep-th/08052601].
}

\lref\LL{
  E.M. Lifshitz and L.P. Pitaevskii, Statistical Physics Part 2, Pergamon Press, Oxford, 1980.
}

\lref\HerzogHS{
C.~P.~Herzog and D.~T.~Son,
  ``Schwinger-Keldysh propagators from AdS/CFT correspondence,''
   JHEP {\bf 0303}, 046 (2003)
   [arXiv:hep-th/0212072].
}

\lref\Marolf{
D.~Marolf,
  ``States and boundary terms: Subtleties of Lorentzian AdS/CFT,''
  JHEP {\bf 0505}, 042 (2005)
  [arXiv:hep-th/0412032].
}

\lref\SkenderisSVR{
   K.~Skenderis and B.~C.~van Rees,
   ``Real-time gauge/gravity duality,''
   [arXiv:hep-th/0805.0150].
}

\lref\KobayashiEA{
   S.~Kobayashi, D.~Mateos, S.~Matsuura, R.~C.~Myers and R.~M.~Thomson,
  ``Holographic phase transitions at finite baryon density,''
    JHEP {\bf 0702}, 016 (2007)
    [arXiv:hep-th/0611099].
}

\lref\SonSS{
    D.~T.~Son and A.~O.~Starinets,
  ``Minkowski-space correlators in AdS/CFT correspondence: Recipe and applications,''
    JHEP {\bf 0209}, 042 (2002)
   [arXiv:hep-th/0205051].
}

\lref\MMT{
  D.~Mateos, R.~C.~Myers and R.~M.~Thomson,
  ``Thermodynamics of the brane,''
  JHEP {\bf 0705}, 067 (2007)
  [arXiv:hep-th/0701132];
}

\lref\BabingtonVM{
  J.~Babington, J.~Erdmenger, N.~J.~Evans, Z.~Guralnik and I.~Kirsch,
  ``Chiral symmetry breaking and pions in non-supersymmetric gauge /  gravity
  duals,''
  Phys.\ Rev.\  D {\bf 69}, 066007 (2004)
  [arXiv:hep-th/0306018].
}

\lref\CJV{
  T.~Albash, V.~G.~Filev, C.~V.~Johnson and A.~Kundu,
  ``A topology-changing phase transition and the dynamics of flavour,''
  arXiv:hep-th/0605088;
  T.~Albash, V.~G.~Filev, C.~V.~Johnson and A.~Kundu,
  ``Global Currents, Phase Transitions, and Chiral Symmetry Breaking in Large
  $N_c$ Gauge Theory,''
  arXiv:hep-th/0605175;
  V.~G.~Filev, C.~V.~Johnson, R.~C.~Rashkov and K.~S.~Viswanathan,
  ``Flavoured large N gauge theory in an external magnetic field,''
  arXiv:hep-th/0701001.
}

\lref\ErdmengerEA{
  J.~Erdmenger, M.~Kaminski, P.~Kerner and F.~Rust,
  ``Finite baryon and isospin chemical potential in AdS/CFT with flavor,''
   arXiv:hep-th/0807.2663;
  J.~Erdmenger, M.~Kaminski and F.~Rust,
  ``Holographic vector mesons from spectral functions at finite baryon or
  isospin density,''
  Phys.\ Rev.\  D {\bf 77}, 046005 (2008)
  arXiv:hep-th/0710.0334.
}

\lref\MyersMST{
  R.~C.~Myers, A.~O.~Starinets and R.~M.~Thomson,
  ``Holographic spectral functions and diffusion constants for fundamental
  matter,''
  JHEP {\bf 0711}, 091 (2007)
  [arXiv:hep-th/0706.0162].
}

\lref\MateosMMT{
  D.~Mateos, R.~C.~Myers and R.~M.~Thomson,
  ``Holographic viscosity of fundamental matter,''
  Phys.\ Rev.\ Lett.\  {\bf 98}, 101601 (2007)
  [arXiv:hep-th/0610184].
}

\lref\MyersMMT{
  D.~Mateos, R.~C.~Myers and R.~M.~Thomson,
  ``Holographic phase transitions with fundamental matter,''
  Phys.\ Rev.\ Lett.\  {\bf 97}, 091601 (2006)
  [arXiv:hep-th/0605046].
}

\lref\HovdeboKM{
  J.~L.~Hovdebo, M.~Kruczenski, D.~Mateos, R.~C.~Myers and D.~J.~Winters,
  ``Holographic mesons: Adding flavor to the AdS/CFT duality,''
  Int.\ J.\ Mod.\ Phys.\  A {\bf 20} (2005) 3428.
}

\lref\AbrikosovKhalatnikov{
  A.~A.~ Abrikosov and I.~M.~Khalatnikov,
  ``The Theory of a Fermi Liquid (the properties of liquid $^3$He at low temperatures)''
  Rep.\ Prog.\ Phys.\ {\bf 22} (1959) 329-367.
}

\Title{\vbox{\baselineskip12pt
}}
{\vbox{\centerline{Comments on Fermi Liquid from Holography}
\vskip.06in
}}
\centerline{Manuela Kulaxizi${}^a$ and  Andrei Parnachev${}^b$}
\bigskip
\centerline{{\it ${}^a$Institute for Theoretical Physics, University of Amsterdam,}}
\centerline{{\it Valckenierstraat 65, 1018XE Amsterdam, The Netherlands }}
\centerline{{\it ${}^b$C.N.Yang Institute for Theoretical Physics, Department of Physics,}}
\centerline{{\it Stony Brook University, Stony Brook, NY 11794-3840, USA}}
\vskip.1in \vskip.1in \centerline{\bf Abstract}
\noindent
We investigate the signatures of Fermi liquid formation in the $\NN=4$ super Yang-Mills theory
coupled to fundamental hypermultiplet at nonvanishing chemical potential for the global $U(1)$
vector symmetry.
At strong 't Hooft coupling the system can be analyzed in terms of the D7 brane
dynamics in $AdS_5\times S^5$ background.
The phases with vanishing and finite charge density are separated at zero temperature
by a quantum phase transition.
In case of vanishing hypermultiplet mass, Karch, Son and Starinets discovered a
gapless excitation whose speed equals the speed of sound.
We find that this zero sound mode persists to all values of the hypermultiplet mass, and
its speed vanishes at the point of  phase transition.
The value of critical exponent and the ratio of the velocities of zero and first sounds
are consistent with the predictions of Landau Fermi liquid theory at strong coupling.
\vfill

\Date{August 2008}


\newsec{Introduction and summary}

Gauge/string duality is a spin-off of string theory which allows access to the
dynamics of strongly coupled field theories by relating them to string theories
in certain backgrounds.
In principle, this relation provides a unique way of understanding phenomena
at strong coupling which cannot be analyzed by other means.
One of the fundamentally interesting questions (with many practical applications)
concerns the behavior of the Fermi systems at strong coupling.
Here we attempt to analyze the possibility of the formation of Fermi liquid
in a very specific system -- $N_f$ $\NN=2$ fundamental hypermultiplets
coupled to $SU(N_c)$ $\NN=4$ super Yang Mills
theory.
As usual, gauge/string duality methods are applicable  in the regime of
$N\ra\infty$, $N_f$ finite, and large four-dimensional 't Hooft coupling $\lambda$.
Recent work on this system in the strong coupling regime includes
\refs{\KobayashiEA\MM\MMT\MyersMS\MyersMST\MateosMMT\MyersMMT\HovdeboKM\MasSG\ErdmengerEA\CJV\NakamuraNX-\GhorokuRE}.

The hypermultiplet field content includes two Weyl fermions and two complex
scalars.
The global symmetry of the theory is $U(N_f)$, but we will be mostly
concerned with its $U(1)_B$ subgroup\foot{The subscript here stands for ``baryon symmetry''.}
and our results are not going to depend on the value of $N_f$.
We will be interested in the situation where  nonvanishing
chemical potential for $U(1)_B$ is turned on.
In the case of free fermions, such a setup would result in the formation of
a degenerate gas of fermions, whose Fermi energy is equal to the value of
the chemical potential $\mu$.
As the interaction is turned on, possible scenarios include formation
of the Fermi liquid, superfluid, superconductor, or, possibly, some other
exotic phase.

Fermi liquid is a state whose properties can be described in terms of the
dynamics of quasiparticles of Fermi statistics taking place in the narrow
region around Fermi surface.
The momentum and energy of a quasiparticle in the degenerate Fermi liquid at $T=0$
are bounded from above by the Fermi momentum $k_F$ and Fermi energy.
Non-relativistic Fermi liquids are characterized by a number of properties.
Among them are heat capacity proportional to $T$ and the existence of zero sound.
In \KarchFA\ the D3-D7 system with vanishing hypermultiplet mass $m$ has been investigated.
It has been noted that the heat capacity of the system
is proportional to $T^6$ (unlike the normal Fermi liquid).
However, on the basis of  the existence of a gapless excitation
(arising as a massless pole in the density-density correlator) \KarchFA\
argued that a novel kind of quantum liquid might be formed.
This massless excitation is associated with zero sound in  \KarchFA,
and we discuss some issues related to this identification
in this paper.

We generalize the work of \KarchFA\ to the finite value of $m$ and
find that the gapless excitation persists and the dispersion relation
has an interesting dependence on $m$.
In the relativistic regime, $\mu\gg m$, we reproduce the results of \KarchFA\
where the speed of zero sound $u_0$  was found to be equal to the speed of
regular sound.
On the other hand, in the non-relativistic regime,  $\mu-m\ll m$, we find that
$u_0$ is proportional to $(\mu-m)^{1\over2}$, and differs from the value of the first
sound, $u_1$ by an overall multiplicative constant.
Such a behavior is consistent with the prediction of Landau Fermi liquid theory.
In particular, the value of the critical exponent is precisely the expected
$1/2$, while the ratio of  $u_0$ and $u_1$ is expected to go to a constant in
the limit of strong coupling.

The rest of the paper is organized as follows.
In the next section we give a brief review of the relevant parts of Landau's theory
of Fermi liquids.
In Section 3 we review the thermodynamics of the D3-D7 system.
At large value of the 't Hooft coupling the dynamics of  fundamental matter
is captured by the DBI action for D7 branes propagating in the
$AdS_5\times S^5$ background \KarchSH.
At low temperatures, increasing the value of chemical potential results in
a phase transition between the phases with vanishing and non-vanishing
charge density.
In the brane setup this corresponds to the brane embedding intersecting the
horizon becoming thermodynamically  preferred.
In the limit of small temperature, the thermodynamics can be studied
analytically \KarchBR.

In Section 4 we derive equations of motion for the
fluctuating fields on the brane.
We largely follow \KovtunEV\ and  \KarchSH, with new ingredient being the
finite value of the hypermultiplet mass $m$.
We then proceed to analyze the equations in the regime of small frequency and momentum,
find the massless excitation and compute its velocity as a function of
the hypermultiplet mass $m$.
We discuss our results in Section 5.
Some technical results appear in the appendices.

%
%
\newsec{Landau Fermi liquid theory}
In this section we review the basics of Landau Fermi liquid theory
and quote the  results for the velocities of first and zero sound.
We also show that the ratio of
these velocities goes to a constant for generic strong interactions between
the quasiparticles.
Non-relativistic Fermi liquid at zero temperature involves the dynamics
of quasiparticles of Fermi statistics whose dispersion relation is not
necessarily equal to that of free fermions.
The momentum distribution is described by a step function $\theta(|\k|-k_F)$
where $k_F$ is the value of the Fermi momentum, which is related to the particle density
via
\eqn\kfn{ n\equiv {N\over V}={k_F^3\over3\pi^3\hbar^3}  }
The quasiparticle description is, in fact, only valid in the small
vicinity of the Fermi surface.
The change in the total energy of the system due to the change in the
distribution function $\delta n(\k)$ is
\eqn\changee{ \delta E=\int \epsilon(k) \delta n(\k) d\tau + \int f(\k,\k') \delta n(\k) \delta n(\k') d\tau d\tau'  }
where $d\tau=V d^3k/(2\pi\hbar)^3$ is the element of the phase space, $\epsilon(k)$ is the
energy of a quasiparticle with momentum $\k$ (whose absolute value we denote by $k$).
In eq. \changee\ the second term describes interaction of the quasiparticles
which takes place only in the narrow region near the Fermi surface.
It is natural then to assume that in $f(\k,\k')$ the quasiparticle momenta lie on the
Fermi surface. The function $f(\k,\k')$ therefore only depends on the relative angle $\vartheta$.

It is convenient to introduce the Fermi velocity and the effective quasiparticle mass by
\eqn\deffv{  \upsilon_F={\p\epsilon(k)\over \p k}\big{|}_{k=k_F},\qquad m^*={k_F\over \upsilon_F}   }
and to write
\eqn\ftheta{  f(\k,\k')= {k_F m^*\over \pi^2\hbar^3 } F(\vartheta)  }
where $ F(\vartheta)$ is a dimensionless function
which can be expanded in terms of  Legendre polynomials,
\eqn\flegendre{  F(\vartheta)=\sum_l (2l+1)\; F_l\; P_l(\cos\vartheta)  }
In the right hand side of \ftheta\ the prefactor is equal to the density of states on the Fermi surface.
According to \LL, the quasiparticle mass is related to the bare mass via
\eqn\bareqp{  {m^*\over m}=1+{F_1\over3}  }

An interesting property of the Fermi liquid at zero temperature is its compressibility,
\eqn\defcompr{  u^2={\p P\over\p\rho}={N\over m} {\p\mu\over\p N }   }
where $P$ is the pressure, $\rho$ is the density and $N$ is the number of particles.
The value of $u$ also defines the speed of (normal) sound in the
Fermi liquid.
According to \LL, it is given by
\eqn\compfla{  u^2={k_F^2\over3 m m^*}\left(1+F_0\right)=
                    {\upsilon_F^2\over3} \left(1+F_0\right)\left(1+{F_1\over3}\right)   }
where in the second equality we used \deffv\ and \bareqp.
Another interesting property of the Fermi liquid is the existence
of a massless excitation in the limit of zero temperature.
This excitation, called zero sound, corresponds to the deformation of the shape
of Fermi surface.
The speed of zero sound $u_0$ is bounded from below by, and is generally proportional to,
the Fermi velocity $\upsilon_F$.
To compute the ratio $s\equiv u_0/\upsilon_F$ one is instructed to solve
the following integral equation
\eqn\inteqs{  (s-\cos\vartheta)\nu(\vartheta,\varphi)=\cos\vartheta\int F(\vartheta') \nu(\vartheta',\varphi') {d\Omega'\over2\pi}}
In eq. \inteqs\ the integral is over the solid angle and function  $\nu(\vartheta,\varphi)$
parametrizes the displacement of the spherical Fermi surface.


In particular, when $F(\vartheta)$ contains only the zeroth $F_0$ and first $F_1$ harmonic \AbrikosovKhalatnikov\
one has
\eqn\szsll{  {s\over2}\log\left({s+1\over s-1}\right)-1=
           {{1\over 3} F_1+1 \over F_0+{1\over 3}F_0 F_1+F_1 s^2 }}
where $s$ is the ratio between the speed of zero sound and the Fermi velocity\foot{Here zero sound velocity refers to
the velocity of the mode for which $\nu(\vartheta,\varphi)$ is isotropic in the plane perpendicular to its momentum.}.
In the limit of noninteracting quasiparticles, the speed of zero sound is equal to the Fermi velocity.
In other words, $s=1$ and $u_0^2/u^2=3$.
[This result is independent of the particular form of $F(\vartheta)$.]
Another interesting case is the limit of strongly interacting quasiparticles
with $F_0$ and $F_1$ considerably larger than unity.
From \szsll\ we infer  $s^2\simeq {F_0 F_1\over 9}$ which combined with \compfla\ leads to
\eqn\zfratio{{u_0^2\over u^2}\simeq 1}
For generic and large $F_l$ one can argue that the
ratio between the speed of zero and first sound goes to a constant in the limit of strong coupling.
This is because for a generic function $F(\vartheta)\sim F$ the right-hand side of \inteqs\
scales like $F$ and therefore $s\sim F$.
But according to  \compfla\ $u\sim F$ as well, which justifies our assertion.

This argument may fail for some non-generic values of $F_l$.
In particular, when only $F_0$ is non-vanishing and large, eq. \szsll\ implies
$s\approx \sqrt{F_0/3}$.
But in this case \compfla\ again leads to $u_0/u\approx1$.

\newsec{D3--D7 System}

In this section we review the D3-D7 system at zero temperature with nonvanishing  chemical potential.
Consider $N_c$ D3 branes extended along $x^0,\ldots,x^3$ directions and $N_f$ D7 branes
extended along $x^0,\ldots,x^7$ directions.
Separating these branes in the $x^8-x^9$ plane
gives a mass to the hypermultiplet whose degrees of freedom are massless excitations of the
fundamental string stretched between the $D3$ and $D7$ branes.
Taking the near horizon limit, $N_c\ra \infty$, $g_s\ra 0$ with $\lambda$ fixed but large and $N_f\ll N_c$
we obtain the $AdS_5\times S^5$ geometry with $N_c$ units of RR five--form flux
and D7--branes propagating in this background.
Studying the holographic dual of the D3-D7 system
with nonvanishing chemical potential involves analyzing the DBI action for the D7 branes embedded in $AdS_5\times S^5$ with
the electric field flux on their worldvolume turned on.
The value of the chemical potential is equal to the asymptotic value  of the gauge field on the brane.

In the following we
review the results of \KarchBR~ where analytic results
for the zero temperature limit of the D3-D7 system were first obtained.\foot{ Note
that our notations are slightly different from those of \KarchBR.}
The background metric can be written as
\eqn\AdS{  ds^2=\left({\rho\over L}\right)^{2} (-dt^2+dx^{i}dx_{i}) + \left({\rho\over L}\right)^{-2} \left(dr^2+r^2 d\Omega_3^2 +dR^2+R^2 d\phi^2\right)}
where $t$ is time,  $i=1,2,3$ denote the spatial directions along the D3--brane,
and $d\Omega_3^2$ is the metric of a unit three-sphere within $S^5$.
In eq. \AdS\ $\rho$ is the radial coordinate transverse to the D3 branes which is further
expressed as
\eqn\radialeq{\rho^2=r^2+R^2}
In this parametrization  $R$ and $\phi$ are the polar coordinates in the $x^8-x^9$ plane,
and
U(1) symmetry allows one to
set $\phi=const$.
Hence, the  brane embedding is specified by a single function $R(r)$.
The induced metric is then
\eqn\dsevmizero{{ds^2_{D7}}=\left({\rho\over L}\right)^2\left(-dt^2+dx^i dx_i\right) + \left({\rho\over L}\right)^{-2} [\left(1+R'(r)^2\right) dr^2+r^2 d\Omega_3^2] }
where $R'(r)={\p R\over\p r}$.
The DBI action for this configuration is now relatively simple.
Warp factors drop out and we have
\eqn\actionzero {S_{D7}=-\N \int dr r^3 \sqrt{1+R'(r)^2-A_0'(r)^2}  }
where $\N={1\over (2\pi)^4}{1\over 2\lambda} N_c N_f$.
Note that we divided by the volume of $R^{1,3}$ therefore \actionzero\ is actually an action density.
Moreover, we fixed the gauge by choosing $A_r=0$ and rescaled $A_0$ according to:
\eqn\rgfzero{ A_0\ra {2\pi} A_0}
Here and in the rest of the paper we set the string length to one.
Given that \actionzero~ contains only first derivatives of the fields $R(r)$ and $A_0(r)$ there are two conserved charges
\eqn\cc {-r^3 {R'\over\sqrt{1+R'^2-(A_0')^2}}=-c \qquad r^3 {A_0'\over \sqrt{1+R'^2-(A_0')^2}}=d  }
One can re-express $R'(r),A_0'(r)$ in terms of $c$ and $d$
\eqn\finaleq {R'={c\over\sqrt{r^6+d^2-c^2}} \qquad A_0'={d\over\sqrt{r^6+d^2-c^2}} }
The bare quark mass $m$ and chemical potential $\mu$ are related
to the asymptotic values of $R(r\ra \infty)$ and $A_0(r\ra \infty)$ as
\eqn\mqmu{R(r\ra \infty)=\tilde{m}=2\pi m, \qquad A_0(r\ra \infty)=\tilde{\mu}=2\pi \mu}

Possible phases of this system are classified by the values of integration constants $c$ and $d$.
For $c=d=0$,
both $R(r)$ and $A_0(r)$ are constant.
In this phase the condensate and the charge density
vanish and the values of
of $R$ and $A_0$ correspond to the quark mass and the chemical potential respectively.
This is the zero temperature limit of the "Minkowski" embedding.
Solutions of \finaleq\ for all other values of $c$ and $d$ satisfying $d^2-c^2>0$
are the zero temperature limits of the finite temperature black hole embeddings.
They are characterized by the boundary conditions
\eqn\bcbhzero{ R(r=0)=0 \qquad A_0(r=0)=0}
and nonvanishing values for both the condensate and the charge density.
Equation \finaleq~ in this case yields
\eqn\RAsol{\eqalign{R(r)&={1\over 6} c \left(d^2-c^2\right)^{-1/3} \BB\left[{r^6\over r^6+d^2-c^2};{1\over 6},{1\over 3}\right]\cr
A_0(r)&={1\over 6} d \left(d^2-c^2\right)^{-1/3} \BB\left[{r^6\over r^6+d^2-c^2};{1\over 6},{1\over 3}\right]}}
where $\BB\left[{r^6\over r^6+d^2-c^2};{1\over 6},{1\over 3}\right]$ denotes the incomplete beta function while
the values of $c$ and $d$  are related to the physical variables $m={\tm\over2\pi}$ and $\mu={\tmu\over 2\pi}$ through
\eqn\mmu{c=\gamma \tm (\tmu^2-\tm^2) \qquad d= \gamma \tmu (\tmu^2-\tm^2) \qquad \gamma= \left({1\over 6} \BB\left[{1\over 6},{1\over 3}\right] \right)^{-3}}
Eq. \mmu\ can be inverted to yield
\eqn\mmucd{\tm=c \gamma^{-{1\over3}} (d^2-c^2)^{-{1\over 3}} \qquad \tmu=d \gamma^{-{1\over 3}} (d^2-c^2)^{-{1\over 3}}\qquad k_F^0 \equiv\tmu^2-\tm^2=\gamma^{-{2\over 3}} (d^2-c^2)^{{1\over 3}}}
Combined with $d^2-c^2>0$, \mmucd~ shows that black hole
embeddings are only realized when the chemical potential is greater than the bare quark mass, $\mu>m$.


The phase structure of the D3-D7 system at $T=0$ has been analyzed in \KarchBR\
with the following results.
When $\mu< m$ there is only one configuration available, the Minkowski embedding.
In this phase, the system is characterized by zero condensate and charge density.
For $\mu>m$  the configuration with brane falling to $R=r=0$, the black hole embedding,
is thermodynamically preferred.
The two phases are separated by a quantum phase transition at $\mu=m$.
The black hole embeddings are
distinguished by the non-vanishing values of both the
condensate and the charge density.
\eqn\cdvalues{ \langle\OO_m\rangle= 2\pi c \N \qquad
                \langle J_0\rangle=2\pi d \N        }
where $c$ and $d$ are related to $m$ and $\mu$ via \mmu.
Note that the expectation values in eq.\cdvalues\ and, in particular,
the charge density, vanish at the point of phase transition $\mu=m$.
In the rest of the paper we will be concerned with the black hole embeddings.

%
%

%
%
%
\newsec{Zero sound}
%
%
We will be  interested in finding the massless excitation
in the D3-D7 system at strong coupling.
This requires analyzing linearized equations of motion which follow from
the DBI action for the D7 brane propagating in $AdS_5\times S^5$.
In this Section we only quote a few key results.
For technical details the reader is encouraged to consult
Appendix A.

The DBI action can be written as\foot{The full action contains a Chern-Simons term.
It can be shown that it does not contribute to quadratic order in the fluctuations for the field configurations considered below.}
\eqn\DBIdef{S_{D7}\sim \int d\Omega_3 \int d^4x\int dr \sqrt{-\det\;\GG}     }
where $\GG=G+\FF$ is the sum of the induced metric and gauge field strength.
Here we will only be interested in perturbations which are independent of
the coordinates on $S^3$. We therefore expand the D7 brane embedding coordinate
$R=R_0+\xi$, the gauge field $\AA_0=A_0^{(0)}+A_0,\AA_i=A_i$ and
the gauge field strength $\FF_{0r}=F_{0r}^{(0)}+F_{0r},\FF_{ij}=F_{ij}$,
where $i,j=1,2,3$ and the superscript $(0)$ denotes the background values of the corresponding fields
(whose profile follows from the results of  the previous section.)
In addition, we set $\AA_{\theta}=0, \forall~\theta\in S^3$ and
choose the $\AA_r=0$ gauge.
To expand the DBI action to quadratic order in fluctuations
one can make use of the formula
\eqn\trdet{ \sqrt{\det\; \GG}=\sqrt{\det \GG_0} \left(
           1+{1\over8}\left(\tr \GG_0^{-1}\delta\GG\right)^2-
                {1\over4}\tr(\GG_0^{-1}\delta\GG\GG_0^{-1}\delta\GG)
    \right)   }
where we write $\GG=\GG_0+\delta\GG$ and $\delta\GG$ stands for the fluctuations.
In \trdet\ there is of course a linear term, but it vanishes by equations of
motion.
It is convenient to keep the fluctuations of the brane embedding $\x$
(but not derivatives of $\x$) as a part of $\GG_0$.
Both the leading term and a term linear in fluctuations do not contain $\xi$,
which means it does not appear at quadratic order in fluctuations.

The action for fluctuations is given by [see also eq. (A.1)]
\eqn\DBI{\eqalign{&S_{D7, fl}=-{1\over 2} \N  \int d^4x \int dr g(r) \left[ \sum_{i} F_{ir}^2
              - f_1(r) F_{0r}^2-L^4 f_2(r) \sum_{i} F_{0i}^2+ L^4 f_3(r)\sum_{i<j} F_{ij}^2+ \right.  \cr
                &\left. +f_5(r) (\p_r\xi)^2-{L^4\over\rho_0^4} (\p_0\xi)^2+L^4 f_4(r) \sum_{i}(\p_i\xi)^2
                -2 L^4 f_6 \sum_{i} F_{0i}(\p_i\xi)-2 f_7(r) F_{0r} (\p_r\xi)\right] }}
where the functions $f_i,\,i=1,\ldots,7$ can be found in (A.2) and (A.3).
It is convenient to use the momentum space representation,
\eqn\mom{A_M(x^{\mu},r)=\int {d^4k\over (2\pi)^4} e^{i k_{\mu}x^{\mu}} \Af_M(k^\mu,r) }
and
\eqn\momxi{\x(x^{\mu},r)=\int {d^4k\over (2\pi)^4} e^{i k_{\mu}x^{\mu}} {\tilde \x}(k^\mu,r) }
We can also choose the direction of wave propagation by
setting  $k_{\mu}=(-\omega,0,0,q)$.
It is important to work with the gauge invariant combination,
\eqn\gaugeinvarinate{E=\tq\Af_0+\tom \Af_{\parallel}}
We can use  Gauss' law (equation of motion for $A_r$),
\eqn\gauss{\tq\Af'_{\parallel}+\tom f_1 \Af'_{0}-\tom f_7 {\tilde \x}'=0}
to express $A_0'$ (or $A_{||}'$) in terms of $E'$ and ${\tilde \x}'$.
Here prime denotes differentiation with respect to $r$.
In these equations and in the following we define $\tq= q L^2=q\sqrt{2\lambda}$ and $\tom=\omega L^2=\omega\sqrt{2\lambda}$.
Equations of motion are in general quite complicated.
In particular, the equations for the longitudinal components of the gauge field
are coupled to the equations for the embedding fluctuation $\xi$.
It turns out, however, that they can be solved analytically
in the two overlapping regimes, much like in \KarchFA.

As explained in Appendix A, in the vicinity of the horizon
equations for $E$ and $\tilde\x$ decouple:
\eqn\eqh{\ddot{E}+{2\over z}\dot{E}+\Omega^2 E=0}
\eqn\eqhx{\ddot{\tilde \x}+{2\over z}\dot{\tilde \x}+\Omega^2 {\tilde \x}=0}
where $\Omega^2\equiv\tom^2 {d^2-c^2\over d^2}$ and the dot denotes differentiation
with respect to $z=1/r$.
The solutions of \eqh\ and \eqhx\ are
\eqn\eqhsol{E(z)=A {{e^{+ i \Omega z}}\over z},\qquad {\tilde \x}(z)=B {{e^{+ i \Omega z}}\over z}}
The positive exponent is singled out by incoming boundary conditions
at the horizon  \refs{\SonSS\HerzogHS\Marolf-\SkenderisSVR},\KarchFA.
In the limit of small $\tom$, $\Omega z\ll 1$ \eqhsol\ becomes
\eqn\eqhsoltwo{  E(z)=i\Omega A+{A\over z},\qquad  {\tilde \x}=i\Omega B+{B\over z} }
On the other hand, for sufficiently small $\tom$ and $\tq$
equations of motion simplify again:
\eqn\eqwzsmallamt{\eqalign{&\dot{Z}+\left({2\over z}+{1\over h f_5}{\tq^2 \dot{f}_1\over \tq^2-\tom^2 f_1}\right) Z-
                         \left({\dot{f}_7\over h f_5}{\tq^2-\tom^2\over \tq^2-\tom^2 f_1}\right) Y=0\cr
                         &\dot{Y}+\left({2\over z}+{\dot{f}_5\over h f_5}{\tq^2-\tom^2\over \tq^2-\tom^2 f_1}\right) Y+
                         \left({\dot{f}_7\over h f_5}{\tq^2\over \tq^2-\tom^2 f_1}\right) Z=0 }}
where $Z=g(z)\dot{E}(z)$, $Y=\tq g(z)\dot{\tilde \x}(z)$, and $h(z)=(1+(d^2-c^2) z^6)/(1-c^2 z^6)$.
 The solution of \eqwzsmallamt\ is given by
\eqn\solYZmt{\eqalign{\dot{E}&=-{z\over\left(1+(d^2-c^2) z^6\right)^{3\over2}}
        \left[ C_1 [ (1-c^2 z^6) \tq^2 -(1+(d^2-c^2) z^6) \tom^2]+C_2 c d z^6 \tq^2\right] \cr
            \dot{\tilde\x}&={q z\over\left(1+(d^2-c^2) z^6\right)^{3\over2}} \left[ C_1 c d z^6-C_2 (1+d^2 z^6)\right]
        }}
where $C_1$ and $C_2$ are arbitrary integration constants.
Eqs. \solYZmt\ can also be obtained directly from eq. (A.6)
by neglecting all non-derivative terms, performing trivial integration,
and using Gauss' law to express ${\tilde A}_0'$ in terms of $E'$ and ${\tilde \x}'$.
Eqs.  \solYZmt\  can in turn be integrated to yield
\eqn\exiinteg{\eqalign{ E&=C_0+\int_0^z {-x dx\over\left(1+(d^2-c^2) x^6\right)^{3\over2}}
        \left[ C_1 [ (1-c^2 x^6) \tq^2 -(1+(d^2-c^2) x^6) \tom^2]+C_2 c d x^6 \tq^2\right] \cr
{\tilde\x}&={\tilde C_0}+\int_0^z
              {q x dx\over\left(1+(d^2-c^2) x^6\right)^{3\over2}} \left[ C_1 c d x^6-C_2 (1+d^2 x^6)\right] } }
The integrals can be expressed in terms of hypergeometric
functions.
We do not need this representation, since only small and large $z$ asymptotics
are important for our purposes.
In particular, near the boundary
\eqn\exibndy{  E= C_0+\OO(z^2),\qquad \xi={\tilde C}_0+\OO(z^2)  }
This implies that the spectrum of quasinormal modes can be obtained by
requiring $C_0={\tilde C}_0=0$.

In the region $z\gg1$ the leading asymptotic behavior of
\exiinteg\ is a constant whose value we can infer by performing integration
from $z=0$ to $z=\infty$.
The subleading $1/z$ term can be  extracted from the expression
for the derivatives \solYZmt.
The results are
\eqn\eqezlarge{ E=C_0+b_1 C_1+b_2 C_2+{a_1 C_1\over z}+{a_2 C_2\over z}   }
where
\eqn\defae{  a_1=-{c^2 \tq^2+(d^2-c^2) \tom^2\over (d^2-c^2)^{3\over2}},\qquad
             a_2={c d \tq^2\over (d^2-c^2)^{3\over2}}  }
and
\eqn\defbe{ b_1={\Gamma\left({7\over6}\right)\Gamma\left({4\over3}\right)\over (d^2-c^2)^{4\over3}\sqrt\pi}
                \left[ (3 c^2-d^2) \tq^2 +3 (d^2-c^2) \tom^2\right],\quad
            b_2= -2 c d \tq^2 {\Gamma\left({7\over6}\right)\Gamma\left({4\over3}\right)\over (d^2-c^2)^{4\over3}\sqrt\pi}  }
Likewise,
\eqn\eqxzlarge{{\tilde\xi}  =  \C_0 +\tb_1 C_1+\tb_2 C_2+ {\ta_1 C_1\over z}+{\ta_2 C_2\over z}     }
where
\eqn\defta{\ta_1=-{c d\over (d^2-c^2)^{3\over2}},\qquad \ta_2={d^2\over (d^2-c^2)^{3\over2}}  }
and
\eqn\deftb{\tb_1={2 c d\Gamma\left({7\over6}\right)\Gamma\left({4\over3}\right)\over (d^2-c^2)^{4\over3}\sqrt\pi},\qquad
            \tb_2=- {\Gamma\left({7\over6}\right)\Gamma\left({4\over3}\right)\over (d^2-c^2)^{4\over3}\sqrt\pi}
                            (3 d^2-c^2)
}
We will focus on the quasinormal mode with the linear dispersion relation
in the regime of small $\tom$ and $\tq$.
In this case we can match the near horizon solutions \eqhsoltwo\ to \eqezlarge\
and \eqxzlarge.
Requiring $C_0=\C_0=0$ and neglecting terms of higher order in $\tom$ and $\tq$
we obtain
\eqn\spectrm{  b_1 C_1+b_2 C_2=0,\qquad \tb_1 \C_1 +\tb_2 \C_2=0  }
These equations can be simultaneously solved with nonvanishing $C_1$ and $C_2$
whenever $b_1 \tb_2-b_2\tb_1=0$.
Using \defbe\ and \deftb\ we arrive at
\eqn\dispr{  \tom^2=u_0^2 \tq^2,\qquad  u_0^2={d^2-c^2\over 3d^2-c^2 }  }
Using \mmu\ we can write the expression for the speed of zero sound as
\eqn\szsmmu{  u_0^2={\mu^2-m^2\over 3\mu^2-m^2}   }
This is the main result of this paper.
In the relativistic limit $\mu\gg m$ we recover the result of \KarchFA, $u_0^2=1/3$.
In the vicinity of the phase transition, where $\delta\mu=\mu-m\ll m$,
the speed of zero sound vanishes as $u_0^2 \sim \delta\mu/m$.
As discussed below, precisely such a behavior is expected from
the Landau's theory of Fermi liquids at strong coupling.

\newsec{Discussion}

Before discussing the significance of \szsmmu\ let us make a comment regarding the identification of the
gapless mode with zero sound.
At first sight, our calculations  imply that the dispersion relation is modified by
a term of the type $-i q^2$.
In particular, in the simpler case of vanishing mass considered in  \KarchFA,
the dispersion relation looks like
\eqn\dispvm{  \tom=\pm{\tq\over 3} -i {\tq^2\over d^{1\over3}} }
Note that in \KarchFA\ the $-i q^2$ behavior of imaginary part of the pole was
important for identification of the massless mode with zero sound.
In the derivation of dispersion relation
it has been assumed that $\tom z\ll 1$ and $d^{1\over3}z\gg 1$ regions overlap.
More precisely, the solutions  \eqh,\ \eqhx\ and \solYZmt\ are
valid up to terms $\OO(d^{-2}z^{-6})$ and  $\OO(w^2z^2)$ respectively\foot{
We thank Andreas Karch for very useful discussions on this issue.}.
This seems to imply that \dispvm\ is only corrected at higher order in $q$.

A number of quantities in the Fermi liquid setup can be computed if the
particle density is known as a function of $\delta\mu=\mu-m$.
In principle, we can compute the charge density using
\eqn\bcd{  \rho= \NN 2\pi d,\qquad \NN={N_f N_c\over (2\pi)^4 2\lambda}  }
It would be equal to the quark density $n$, if the quarks were the
only degrees of freedom.
This would allow one to determine $k_F$ as a function of $\mu$.
However the situation is more complicated, as
both the fermions and the scalars in the fundamental hypermultiplet
are charged under $U(1)_B$.
At vanishing coupling the system is unstable due to the condensation of bosons.
However the lagrangian contains a term which is quartic in scalar
fields.
Hence, at large $\lambda$ the condensation of bosons can be stabilized,
and indeed the solution for the brane embedding does not exhibit
instabilities.
At large coupling one can use the holographic dictionary to compute the
expectation value of the operator which contains the scalar bilinear.
And, as noted in \KarchBR, the dependence of this condensate on $\delta\mu$ is consistent with
the quartic potential for the scalars and a quadratic term of the form $(m^2-\mu^2)q q^*$.
In summary, using \bcd\ to determine $k_F$ would  yield $k_F\sim (d/\lambda)^{1\over3}\sim (m^2\delta\mu/\lambda)^{1\over3}$
but this result is unreliable due to contribution of the boson condensate to the charge
density.
However the expression for $\rho$ can still serve as an upper bound on $n$.

We can deduce the critical behavior of various quantities using
the results reviewed in Section 2.
Writing  \defcompr\  in the form
\eqn\undmu{   u^2={n\over m} {\p\delta\mu\over\p n}  }
and noting that $n$ vanishes near the critical point faster than
$\delta\mu$, we infer $u^2\sim\delta\mu$
and, consecutively, $\upsilon_F\sim\delta\mu^{1\over2}$.
This implies a dispersion relation of the form $\epsilon(k)=k^2/2m^*+\OO(k^3)$.
Hence, in the vicinity of the critical point the density of quarks behaves like
$n\sim k_F^3\sim \delta\mu^{3\over2}$.
This, in turn, leads to
\eqn\uapp{  u^2\approx {2 \delta\mu \over 3 m}   }

As reviewed in Section 2, the speed of zero sound is proportional to Fermi
velocity as well, and must therefore vanish at the critical point as $u_0\sim \delta\mu^{1\over2}$.
This is precisely what eq. \szsmmu\ implies in the limit $\delta\mu\ll m$:
\eqn\unotapp{  u_0^2\approx {\delta\mu \over m}   }
Hence, we observe that our string theoretic computation reproduces
(in a rather nontrivial fashion) the critical exponent predicted
by the phenomenological theory.

What about the coefficient in \unotapp?
As reviewed in Section 2, in the limit of noninteracting fermions
$u^{(0)}_0=\upsilon_F$ or, in other words, $(u^{(0)}_0)^2=2\delta\mu/m$.
This implies $u^{(0)}_0/u^{(0)}=\sqrt{3}$.
However this ratio is expected to be modified when the coupling
is strong.
Indeed, our result $u_0/u=\sqrt{3/2}$ differs from the value
at weak coupling by a factor of $\sqrt{2}$.
This is not surprising since, as discussed in Section 2,
the ratio of the velocities of first and zero
sound goes to a constant which is $\OO(1)$ for
generic strong interaction between quasiparticles.

\bigskip
\noindent
{\bf Acknowledgements:}
We thank  E. Pomoni, L. Rastelli, M. Taylor and especially A. Starinets for discussions and comments on
the manuscript. We also thank the organizers of the Simons Workshop in Mathematics and Physics
where part of this work was completed. M.K acknowledges support from NWO Spinoza grant.
A.P. is grateful to the organizers of Monsoon workshop in String Theory and TIFR, Mumbai,
for hospitality.

%
%
%
%
%
%
%
%
%
%
%
%
%
%

\appendix{A}{Fluctuations}

The action is
\eqn\actionDBI{\eqalign{&S_{D7, fl}=-{1\over 2} \N  \int d^4x \int dr g(r) \left[ \sum_{i} F_{ir}^2
              - f_1(r) F_{0r}^2-L^4 f_2(r) \sum_{i} F_{0i}^2+ L^4 f_3(r)\sum_{i<j} F_{ij}^2+ \right.  \cr
                &\left. +f_5(r) (\p_r\xi)^2-{L^4\over\rho_0^4} (\p_0\xi)^2+L^4 f_4(r) \sum_{i}(\p_i\xi)^2
                -2 L^4 f_6(r) \sum_{i} F_{0i}(\p_i\xi)-2 f_7(r) F_{0r} (\p_r\xi)\right] }}
where the functions $f_i$ with $i=1,\cdots 7$ are defined in terms of the background fields as follows
\eqn\deffunactions{\eqalign{g(r)&={r^3\over \sqrt{1+R'^2-F_{0r}^{(0)2}}} \qquad
                            f_1(r)={1+R'^2\over 1+R'^2-F_{0r}^{(0)2}} \qquad
                           f_2(r)={1+R'^2\over \rho_0^4} \cr
                           f_3(r)&= {1+R'^2-F_{0r}^{(0)2}\over\rho_0^4 } \qquad
                           f_4(r)={1-F_{0r}^{(0)2}\over \rho_0^4}\qquad
                           f_5(r)={1-F_{0r}^{(0)2}\over 1+R'^2-F_{0r}^{(0)2}}\cr
                           &\qquad f_6(r)=-{R'_0 F_{0r}^{(0)}\over \rho_0^4}\qquad\qquad
                           f_7(r)=-{R'_0 F_{0r}^{(0)}\over 1+R'^2-F_{0r}^{(0)2}}
                            }}
Using \finaleq~ we can rewrite these functions as
\eqn\defbg{\eqalign{g(r)&=\sqrt{r^6+d^2-c^2} \qquad f_1(r)=1+{d^2\over r^6} \qquad f_2(r)={r^6+d^2 \over r^6+d^2-c^2}{1\over \rho_0^4} \cr
                            f_3(r)&={r^6\over r^6+d^2-c^2}{1\over \rho_0^4}\qquad f_4(r)={r^6-c^2\over r^6+d^2-c^2} {1\over\rho_0^4} \qquad
                            f_5(r)=1-{c^2\over r^6} \cr
                            &\qquad f_6(r)={c d\over r^6+d^2-c^2}{1\over \rho_0^4}  \qquad\qquad f_7(r)={c d\over r^6}
                             }}
with $\rho_0(r)$ given by \radialeq. It is useful to note that not all of these functions are independent from one
another. For instance
\eqn\relbtwnfs{f_2=f_3 f_1\qquad f_6=f_3 f_7 \qquad f_4=f_3 f_5}
These identities will be particularly useful in expressing the equations of motion in a compact manner.
The field equations are more conveniently expressed in the momentum space representation
\eqn\momsra{A_M(x^{\mu},r)=\int {d^4k\over (2\pi)^4} e^{i k_{\mu}x^{\mu}} \Af_M(k^\mu,r) }
and
\eqn\momsrxi{\x(x^{\mu},r)=\int {d^4k\over (2\pi)^4} e^{i k_{\mu}x^{\mu}} {\tilde \x}(k^\mu,r) }
Further choosing $k_\mu=\left(-\omega,0,0,q\right)$ we find
\eqn\equationcs{\eqalign{&\p_r\left[g f_1 (\p_r \Af_0)- g f_7 (\p_r \xf)\right]-\tom\tq g f_2 \Af_{\parallel}-\tq^2 g f_2 \Af_0+\tq^2 g f_6 \xf=0\cr
                         &\p_r\left[g(\p_r\Af_{\parallel})\right]+\tom\tq g f_2 \Af_0+\tom^2 g f_2 \Af_{\parallel}-\tom\tq g f_6 \xf=0\cr
                         &\p_r\left[g f_5(\p_r\xf)+g f_7 (\p_r\Af_0)\right]+\tom^2 {g\over\rho_0^4}\xf-\tq^2 g f_4 \xf- \tq\tom g f_6 \Af_{\parallel}-
                             \tq^2 g f_6 \Af_0=0\cr
                         &\p_r [g(\p_r \Af_{\perp})]+ g[\tom^2 f_2-\tq^2 f_3] \Af_{\perp}=0 }}
Here we have absorbed the 't Hooft coupling constant $\lambda$ into the variables $\tom$,$\tq$
defined as $\tq= q L^2=q\sqrt{2\lambda}$ and $\tom=\omega L^2=\omega\sqrt{2\lambda}$.
To fix the residual gauge invariance we impose Gauss law
\eqn\gausseq{\tq\Af'_{\parallel}+\tom f_1\Af'_0-\tom f_7 \xf'=0}
Choosing the gauge invariant combination
\eqn\ginvcmb{E=\tq\Af_0+\tom \Af_{\parallel}}
and using \gausseq\ to express the first derivative of $\Af_0$ in terms of $E'$ and $\xf'$
\eqn\eprime{\Af'_0={\tq\over \tq^2-\tom^2 f_1} \left(E'-{\tom^2\over\tq} f_7 \xf'\right)}
leads to
\eqn\eqcoupled{\eqalign{&E''+\left({g'\over g}+{f'_1\over f_1}+ {\tom^2 f'_1\over \tq^2 -\tom^2 f_1}\right)E'+f_3(\tom^2 f_1-\tq^2)E- \cr
                        &-{f_7\over f_1} \left[X''+\left({g'\over g}+{f'_7\over f_7}+{\tom^2 f'_1\over \tq^2 -\tom^2 f_1}\right)X'+
                        f_3(\tom^2 f_1-\tq^2)X\right]=0  \cr
                        &{f_7\over f_5}{\tq^2\over \tq^2-h\tom^2} \left[E''+\left({g'\over g}+{f'_7\over f_7}+
                         {\tom^2 f'_1\over \tq^2 -\tom^2 f_1}\right)E'+f_3(\tom^2 f_1-\tq^2)\right] E+ \cr
                        &+ X''+\left({g'\over g}+{f'_5\over f_5}+ {\tom^2 f'_1\over \tq^2 -\tom^2 f_1}-{\tom^2 h'\over \tq^2 -\tom^2 h}\right)X'+
                        f_3(\tom^2 f_1-\tq^2)X=0
                         }}
where
\eqn\functionsdef{h\equiv {1\over f_4\rho_0^4}={f_7^2+f_1 f_5\over f_5}={r^6+(d^2-c^2)\over r^6-c^2}}
and we defined $X=q \xf$.
Performing a change of variable from $r$ to $z={1\over r}$ \eqcoupled\ is recast into
\eqn\eqcoupledz{\eqalign{&\ddot{E}+\left({2\over z}+{\dot{g}\over g}+{\dot{f}_1\over f_1}+
                        {\tom^2 \dot{f}_1\over \tq^2-\tom^2 f_1}\right)\dot{E}+{f_3\over z^4}(\tom^2 f_1-\tq^2)E- \cr
                        &-{f_7\over f_1} \left[\ddot{X}+\left({2\over z}+{\dot{g}\over g}+{\dot{f}_7\over f_7}+{\tom^2 \dot{f}_1\over \tq^2 -
                        \tom^2 f_1}\right)\dot{X}+
                        {f_3\over z^4}(\tom^2 f_1-\tq^2)X\right]=0  \cr
                        &{f_7\over f_5}{\tq^2\over \tq^2-h\tom^2} \left[\ddot{E}+\left({2\over z}+{\dot{g}\over g}+{\dot{f}_7\over f_7}+
                         {\tom^2 \dot{f}_1\over \tq^2 -\tom^2 f_1}\right)\dot{E}+{f_3\over z^4}(\tom^2 f_1-\tq^2)E \right] + \cr
                        &+ \ddot{X}+\left({2\over z}+{\dot{g}\over g}+{\dot{f}_5\over f_5}+ {\tom^2 \dot{f}_1\over \tq^2 -\tom^2 f_1}-
                        {\tom^2 \dot{h}\over \tq^2 -\tom^2 h}\right)\dot{X}+
                        {f_3\over z^4}(\tom^2 f_1-\tq^2)X=0
                            }}
where dots indicated differentiation with respect to the variable $z$.
Observe that \eqcoupledz\ reduces to two independent equations for $E$ and $X$ near the boundary
since the prefactors
\eqn\factorsbnd{{f_7\over f_1}={c d z^6\over 1+d^2 z^6}\qquad {f_7\over f_5}{\tq^2\over \tq^2-h\tom^2}\simeq {c d z^6\over 1-c^2 z^6}{\tq^2\over\tq^2-\tom^2}}
vanish when $z\rightarrow 0$. We are then left with
\eqn\eqcoupledbnd{\ddot{E}(z)-{1\over z}\dot{E}(z)+\left(\tom^2-\tq^2\right)E(z)=0\qquad \ddot{X}(z)-{1\over z} \dot{X}(z)+\left(\tom^2-\tq^2\right)X(z)=0}
The general solution of this equation can be written in terms of Bessel functions of the first kind as
\eqn\eqcoupledbsol{\eqalign{E(z)&=\AA_1 Y_1\left[z \sqrt{\tom^2-\tq^2}\right]+\BB_1 J_1\left[z \sqrt{\tom^2-\tq^2}\right]\cr
                        X(z)&=\AA_2 Y_1\left[z \sqrt{\tom^2-\tq^2}\right]+\BB_2 J_1\left[z \sqrt{\tom^2-\tq^2}\right]}}
Expanding around $z=0$ yields
\eqn\eqbcompare{\eqalign{E(z)&\simeq \AA_1 \left[1+ {1\over 4} \left(\tom^2-\tq^2\right)\left(1-2\tilde{\gamma}-
                            \ln{\left[{\tom^2-\tq^2\over 4} z^2\right]}\right) z^2\right]+\BB_1 z^2\cr
                          X(z)&\simeq \AA_2 \left[1+ {1\over 4} \left(\tom^2-\tq^2\right)\left(1-2\tilde{\gamma}-
                            \ln{\left[{\tom^2-\tq^2\over 4} z^2\right]}\right) z^2\right]+\BB_2 z^2
                         }}
This identifies the constants $\AA_i$ and $\BB_i$ for $i=1,2$ with the coefficients of the non-normalizable and
the normalizable mode respectively.

In the vicinity of the horizon on the other hand notice that
\eqn\functionscoupled{\eqalign{&{\dot{g}\over g}\simeq -{3\over z}{1\over (d^2-c^2)z^6} \qquad {\dot{f}_1\over f_1}\simeq {6\over z}\qquad
                        {\tom^2\dot{f}_1\over \tq^2-\tom^2 f_1}\simeq -{6\over z} \cr
                       & {\dot{f}_5\over f_5}\simeq {6\over z}\qquad
                         {\dot{f}_7\over f_7}= {6\over z}\qquad
                      {\tom^2 \dot{h}\over\tq^2-\tom^2 h}\simeq {6\over z}{d^2 \over c^4 z^6} {\tom^2\over \tq^2+{d^2-c^2\over c^2} \tom^2}}}
whereas
\eqn\constterm{{f_3\over z^4}\left(\tom^2 f_1-\tq^2\right)\simeq \tom^2 {d^2-c^2\over d^2}}
since $R_0(z)$ behaves for large $z$ as ${c\over\sqrt{d^2-c^2}}{1\over z}$ .
It follows that eq. \eqcoupledz\ reduces to
\eqn\eqcoupledzhor{\eqalign{&\ddot{E}+{2\over z} \dot{E}+\Omega^2 E-{c\over d} \left[\ddot{X}+{2\over z}\dot{X}+\Omega^2 X\right]=0\cr
                        &\ddot{X}+{2\over z}\dot{X}+\Omega^2 X+{d\over c} {\tq^2\over \tq^2+{d^2-c^2\over c^2}\tom^2}
                         \left[\ddot{E}+{2\over z}\dot{E}+\Omega^2 E\right]=0}}
where we defined $\Omega=\tom \sqrt{{d^2-c^2\over d^2}}$.
Multiplying either of the equations with appropriate factors and adding them up yields
the following system of decoupled equations near the 'horizon'
\eqn\eqhorboth{\ddot{E}+{2\over z} \dot{E}+\Omega^2 E=0\qquad \ddot{X}+{2\over z}\dot{X}+\Omega^2 X=0}
with solutions
\eqn\horsol{E(z)=A {e^{i\Omega z}\over z}\qquad X(z)=B {e^{i\Omega z}\over z}}
Note that the incoming wave boundary condition singled out the positive exponent in \horsol .

Eq. \eqcoupledz\ can be simplified even further.
\eqn\decoupledalm{\eqalign{& \ddot{E}+\left({2\over z}+{\dot{g}\over g}+{1\over h f_5}{\tq^2 \dot{f}_1\over \tq^2-\tom^2 f_1}\right)\dot{E}+
                          {f_3\over z^4}\left(\tom^2 f_1-\tq^2\right) E-
                          \left({\dot{f}_7\over h f_5}{\tq^2-\tom^2\over \tq^2-\tom^2 f_1}\right)\dot{X}=0\cr
                           &\ddot{X}+\left({2\over z}+{\dot{g}\over g}+{\dot{f}_5\over h f_5}{\tq^2-\tom^2\over \tq^2-\tom^2 f_1}\right)\dot{X}+
                           {f_3\over z^4}\left(\tom^2 f_1-\tq^2\right) X+\left({\dot{f}_7\over h f_5}{\tq^2\over \tq^2-\tom^2 f_1}\right)\dot{E}=0
                            }}
To arrive at \decoupledalm\ multiply either of the equations with the relevant factor and add it to the other one.
Note that the equations for $E$ and $X$ are now coupled only through first derivative terms of the fields.
This form is particularly useful when considering the region $\Omega z\ll 1$. In this case, terms without
derivatives of the fields $E$ and $X$ can be neglected to yield:
\eqn\eqwzsmalla{\eqalign{&\dot{Z}+\left({2\over z}+{1\over h f_5}{\tq^2 \dot{f}_1\over \tq^2-\tom^2 f_1}\right) Z-
                         \left({\dot{f}_7\over h f_5}{\tq^2-\tom^2\over \tq^2-\tom^2 f_1}\right) Y=0\cr
                         &\dot{Y}+\left({2\over z}+{\dot{f}_5\over h f_5}{\tq^2-\tom^2\over \tq^2-\tom^2 f_1}\right) Y+
                         \left({\dot{f}_7\over h f_5}{\tq^2\over \tq^2-\tom^2 f_1}\right) Z=0 }}
where $Z=g(z)\dot{E}(z)$ and $Y=g(z)\dot{X}(z)$. Solutions for $Z$ and $Y$ are given by
\eqn\solYZ{\eqalign{Z&={d^2 \hat{C}_1
                      \left[\tq^2-\tom^2 \left(1+(d^2-c^2)z^6\right)\right]+\tq^2 \hat{C}_2 \left[-\tq^2 \left(1-c^2 z^6\right)+\tom^2 \left(1+(d^2-c^2)z^6\right) \right] \over c d \tq^2 z^2 \left[1+(d^2-c^2)z^6\right] }\cr
            Y&={1\over z^2}{\hat{C}_1\over 1+(d^2-c^2)z^6}+ z^4 {\hat{C}_2\over  1+(d^2-c^2)z^6}
                                }}
where the integration constants $\hat{C}_1$ and $\hat{C}_2$ are related to the ones appearing in the main text through
\eqn\constantrel{\hat{C}_1=-C_2 \tq^2 \qquad \hat{C}_2=-d\left(d C_2-c C_1\right)\tq^2}
Eq. \solYZ\ can be easily integrated to yield an expression for $E$ and $X$ in terms of hypergeometric
functions.

\appendix{B}{The matching technique in detail.}

Here we investigate in detail the regions of applicability of the matching technique used in section 5.
Let us focus on understanding the precise conditions under which eq. \decoupledalm\ reduces to either
of eqs. \eqhorboth\ or \eqwzsmalla.

Observe that in the vicinity of the horizon, rather when $z$ is large enough so that
$z\gg (d^2-c^2)^{-{1\over 6}}> d^{-{1\over 3}}$ the
following simplifications occur
\eqn\secondterma{\eqalign{ &\qquad\qquad{\dot{g}\over g} \simeq -{3\over z} {1\over (d^2-c^2)z^6} \qquad\qquad
                {1\over h f_5} {\tq^2 \dot{f}_1\over \tq^2-\tom^2 f_1}\simeq
                {6\over z} {d^2\over d^2-c^2} {1\over 1-{\tom^2\over \tq^2} d^2 z^6}\cr
         &\qquad\qquad\qquad {\dot{f}_5\over h f_5} {\tq^2 -\tom^2\over \tq^2-\tom^2 f_1}  \simeq
          -{6\over z}{c^2\over d^2-c^2} {1-{\tom^2\over\tq^2}\over 1-{\tom^2\over \tq^2} d^2 z^6}\cr
                &{\dot{f}_7\over h f_5} {\tq^2 -\tom^2\over \tq^2-\tom^2 f_1} \simeq
              {6\over z}{c d\over d^2-c^2} {1-{\tom^2\over\tq^2}\over 1-{\tom^2\over \tq^2} d^2 z^6}\qquad
               {\dot{f}_7\over h f_5} {\tq^2 \over \tq^2-\tom^2 f_1} \simeq
              {6\over z}{c d\over d^2-c^2} {1\over 1-{\tom^2\over \tq^2} d^2 z^6}
                 }}

Moreover, given that $R(z)$ for large $z$ behaves like $R(z)\simeq {c\over\sqrt{d^2-c^2}} {1\over z}$
\eqn\thirdterma{{f_3\over z^4} (\tom^2 f_1-\tq^2)\simeq {\tq^2 (d^2-c^2)\over d^2} {1\over d^2 z^6} \left({\tom^2\over\tq^2} d^2 z^6-1\right)}
and if we additionally assume that
$z\gg d^{-{1\over 3}} \left({\tq\over\tom}\right)^{{1\over 3}}$ eq. \decoupledalm\ reduces to \eqhorboth.
\eqn\eqhorboth{\ddot{E}+{2\over z} \dot{E}+\Omega^2 E=0\qquad \ddot{X}+{2\over z}\dot{X}+\Omega^2 X=0}
with $\Omega$ defined to be $\Omega^2\equiv\tom^2 {d^2-c^2\over d^2}$.
The solution of \eqhorboth\ is given by
\eqn\horsol{E(z)\sim {e^{i\Omega z}\over z}\qquad X(z)\sim {e^{i\Omega z}\over z}}
When furthermore $z\ll {1\over\Omega}$ the solution becomes
\eqn\eqhsolred{E(z)=A i\Omega+{A\over z} \qquad E(z)=B i\Omega+{B\over z} }
with $A,B$ overall multiplicative constants. In summary, \eqhsolred\ consistently
describes the behavior of the solution of eq. \decoupledalm\ as long as $z$ lies within
the annulus $Max \left[(d^2-c^2)^{-{1\over 6}},d^{-{1\over 3}} \left({\tq\over\tom}\right)^{{1\over 3}}\right]\ll z\ll {1\over \Omega}$.
This in turn implies that we are effectively exploring the region of parameter space in which
\eqn\regionpara{ Max\left[(d^2-c^2)^{-{1\over 6}},d^{-{1\over 3}}
   \left({\tq\over\tom}\right)^{{1\over 3}}\right]\ll {1\over \Omega}}

Now let us investigate the behavior of \decoupledalm\ in the regime $\Omega z\ll 1$ first.
Note that terms without derivatives always appear with the prefactor ${f_3\over z^4} \left(\tom^2 f_1-\tq^2\right)$ as compared to
the two derivative terms. Moreover, this prefactor is given by the difference of two monotonic functions.
Its absolute value is therefore bounded above by the absolute value of their sum.
The latter is a monotonic function which tends to a constant for both large and small $z$.
In particular,
\eqn\htwob{h(z)\equiv \left|{f_3\over z^4} \left(\tom^2 f_1+\tq^2\right)\right|=\left \{
\eqalign{
       & \Omega^2 \qquad\qquad\quad z\gg Max\left[(d^2-c^2)^{-{1\over 6}},\left({\tq\over\tom}\right)^{{1\over 3}} d^{-{1\over 3}}\right] \cr
       &\left|\tom^2+\tq^2\right| \qquad z\ll Min\left[{1\over\tm},(d^2-c^2)^{-{1\over 6}},
         \left({\tq\over\tom}\right)^{{1\over 3}} d^{-{1\over 3}}\right]}
         \right.
}
It follows that terms without derivatives can be neglected as long as they are small compared to the two derivative terms.
This means that for $z\ll Min\left[{1\over\Omega},{1\over \tom} \left|1+{\tq^2\over\tom^2}\right|^{-{1\over 2}}\right]$
eq. \decoupledalm\ consistently reduces to
\eqn\eqwzsmallb{\eqalign{&\dot{Z}+\left({2\over z}+{1\over h f_5}{\tq^2 \dot{f}_1\over \tq^2-\tom^2 f_1}\right) Z-
                         \left({\dot{f}_7\over h f_5}{\tq^2-\tom^2\over \tq^2-\tom^2 f_1}\right) Y=0\cr
                         &\dot{Y}+\left({2\over z}+{\dot{f}_5\over h f_5}{\tq^2-\tom^2\over \tq^2-\tom^2 f_1}\right) Y+
                         \left({\dot{f}_7\over h f_5}{\tq^2\over \tq^2-\tom^2 f_1}\right) Z=0 }}
with $Z=g(z)\dot{E}(z)$ and $Y=g(z)\dot{X}(z)$.

As previously explained, we would like \eqhorboth\ and \eqwzsmalla\ to be simultaneously valid.
This implies that we can investigate \eqcoupledz\ in the following region of parameter space
\eqn\parfinal{Max\left[(d^2-c^2)^{-{1\over 6}},d^{-{1\over 3}} \left({\tq\over\tom}\right)^{{1\over 3}}\right]\ll
              Min\left[{1\over \Omega},{1\over \tom} \left|1+{\tq^2\over\tom^2}\right|^{-{1\over 2}}\right]}
It is manifest from \parfinal\ that the analysis of section 5 covers the region of small $\tom,\tq$.


\footatend\vfill\supereject\immediate\closeout\rfile\writestoppt
\baselineskip=14pt\centerline{{\bf References}}\bigskip{\frenchspacing%
\parindent=20pt\escapechar=` \input refs.tmp\vfill\eject}\nonfrenchspacing
\end